\documentclass[epj,final]{svjour}

\usepackage{latexsym}
\usepackage{url}
\usepackage{amsfonts}
\usepackage{amsmath, amssymb}
\RequirePackage{graphicx, subcaption}

\begin{document}
\title{Cosmic voids and induced hyperbolicity. III. tracing redshift dependence}
\author{M.~Samsonyan\inst{1}, A.A.~Kocharyan\inst{2}, A.~Stepanian\inst{1}, V.G.~Gurzadyan\inst{1,3}% etc
}                     % Do not remove
\institute{Center for Cosmology and Astrophysics, Alikhanian National Laboratory and Yerevan State University, Yerevan, Armenia  \and 
School of Physics and Astronomy, Monash University, Clayton, Australia \and
SIA, Sapienza Universita di Roma, Rome, Italy}
\date{Received: date / Revised version: date}

\abstract{The currently released datasets of the observational surveys reveal the redshift dependence of the physical features of cosmic voids. We study the void induced hyperbolicity, that is the deviation of the photon beams propagating the voids, taking into account the redshift dependence of the void size indicated by the observational surveys. The cumulative image distortion parameter is obtained for the case of a sequence of variable size voids and given underdensity parameters. The derived formulae applied along with those of redshift distortion ones, enable one to trace the number and the physical parameters of the line-of-sight voids from the analysis of the distortion in the galactic surveys.} 
\PACS{
      {98.80.-k}{Cosmology}   %\and
      %{PACS-key}{describing text of that key}
     } % end of PACS codes
%} %end of abstract
%
\maketitle

\section{Introduction}

The role of cosmic voids as of principal features in revealing the large-scale Universe has been steadily increasing due to the advanced observational surveys of the recent decades, \cite{Au,Nad,Pan,Dav,Cor,Se} and references therein. The release of the data of the GIGANTES project \cite{Kr}, as of the most extensive catalog suite containing over 1 billion voids up to the redshift $z=2$, provided a landscape for testing the models for the cosmic web formation. 

Cosmic voids have also become an efficient tool to test fundamental physical principles, modified gravity, dark sector models, e.g. \cite{Amb,Cap,Wil}. The possibility for the Local Group as situated in a local void is discussed e.g. \cite{GC,Cai,Ding}. The void nature of the Cold Spot of the Cosmic Microwave Background sky has been concluded \cite{spot} based on the  estimation of Kolmogorov's stochasticity parameter in the Planck's microwave data for that region, as indication of the void-induced hyperbolicity of the photon beam propagation \cite{GK1,GK2}.

The data  in \cite{Kr} regarding the void distribution and parameter statistic reveal the variation of the void size over the redshift. In the present study we will concentrate on the role of the void size variation in the cumulative image distortion due to the hyperbolicity of the photon beams propagating the cosmic voids. We use the methods developed in the theory of dynamical systems \cite{An,Arn}, include those regarding the hyperbolicity of geodesic flows. Certain features of the void-induced hyperbolicity of photon beams and its observable consequences are studied in \cite{Sam1,Sam2}.

The currently available high redshift data on the cosmic voids increase the interest to the redshift dependence on the void induced observable effects, including their contribution to the image distortion due to the hyperbolicity of the photon beams. Therefore, here we will study the cumulative image distortion taking into account the redshift dependence of the sizes of the line-of-sight voids.

\section{The Jacobi equation}
     
		The hyperbolicity of the geodesic flows defined in Riemmanian manifold can be defined analysing the solutions of the Jacobi equation of the geodesic deviation \cite{An,Arn}
\begin{equation}
\nabla_u \nabla_u n + Riem(n,u)u=0,
\end{equation}
where $n$ is the deviation and the Riemann tensor is
\begin{eqnarray*}
& &
Riem(n,u)u=([\nabla_n,\nabla_u] - \nabla_{[n,u]}) u =
[\nabla_n,\nabla_u] u,\\
& &
\nabla_n u - \nabla_u n = [n,u] = 0.
\end{eqnarray*}
At the orthogonality of the deviation and velocity vectors
\begin{equation}
<n,u>=0,
\end{equation}
the Jacobi equation will turn to
\begin{equation}
\frac {d^2\parallel n \parallel ^2}{ds^2}=
     -2K_{u,n}\parallel n \parallel ^2
    + 2\parallel \nabla_u n \parallel ^2,
\end{equation}
where the two-dimensional curvature is
\begin{equation}
K_{u,n}=
%\frac{<Riem(n,u)u,n>}{\parallel u \parallel ^2\parallel n \parallel ^2 - <u,n>^2} =\\
\frac{<Riem(n,u)u,n>}
{\parallel n \parallel ^2}.
\end{equation}
 
When  $K_{u,n}$ is negative the solutions of the Jacobi equation diverge
not slower than the exponential function
\begin{equation}
\parallel n(s)\parallel \geq \frac{1}{2}
\parallel n(0)\parallel \exp (\sqrt {-2k} s), \qquad s>0
\end{equation}
if the $K_{u,n}$ is negative 
\begin{equation}
k=\max_{u,n} \{K_{u,n}\}<0.
\end{equation}
When this condition is fulfilled the geodesic flows are Anosov systems \cite{An}. 

The null geodesic flows describing the photon beams passing through the voids reveal deviation so that the under-density parameter plays the role of the negative curvature \cite{GK1,GK2}.  Certain properties of the void-induced hyperbolicity have been studied in \cite{Sam1,Sam2}, and here we will reveal the further observable effects.

\section{Hyperbolicity vs distortion}

It has been shown in \cite{GK1}, that for every (d+1)-dimensional Lorentzian manifold, the twice averaged Jacobi equation has the form
\begin{equation}\label{gdev}
\frac{d^2 l}{d \eta^2}+ \frac{\mathfrak{r}_u}{d-1}l =0\,,
\end{equation}
where $\mathfrak{r}_u$ is the d-dimensional spatial Ricci scalar (the details of justification for these averages can be found in \cite{GK1}). On the other hand, for FLRW  metric with small perturbation $\phi$, the line element can be written as ($c=1$)
\begin{equation}\label{FLRW}
ds^2 = -(1+2\phi) dt^2 + (1- 2\phi) a^2(t) d\sigma^2\,, 
\end{equation}
where depending on sign of the sectional curvature  $k$ of the spatial geometry, $d\sigma^2$ represents the spherical, $k=1$, Euclidean, $k=0$ or hyperbolic, $k=-1$, geometries. Meantime, the perturbation field $\phi$, defined over the above metric, satisfies the following conditions \cite{Holz}
\begin{equation}\label{Pert}
|\phi|\ll1, \quad \left(\frac{\partial \phi}{\partial t}\right)^2\ll a^{-2}(t) ||\phi||^2\, .
\end{equation}

The Ricci scalar is 
\begin{align}\label{RicciL}
\frac{\mathfrak{r}_u}{2}
&=k + \frac{16 \pi G}{3}\, \langle\rho_0\rangle a_0^3\ \Theta\ \tilde{\delta},
\end{align}
where 
\begin{equation}
\tilde{\delta}=\frac{\delta\rho}{a\langle\rho\rangle}
=\frac{\rho-\langle\rho\rangle}{a\langle\rho\rangle}
\end{equation}

and 
\begin{equation}\label{theta}
\Theta= \frac{3}{4}\left(1+ \frac{(\nabla^2\phi,\mathbf{u}\otimes\mathbf{u})}{\Delta\phi}\right)\,.
\end{equation}
$\Theta$ reflects the anisotropy of photon beams, $\Theta=1$ corresponding to a spherical distribution i.e. $\langle\mathbf{u}\otimes\mathbf{u}\rangle=\frac{1}{3}\mathbf{\gamma}$.

We can rewrite this using the cosmological parameters \cite{Sam1}
\begin{equation}\label{CosPar}
\Omega_k= - \frac{k}{a^2(t) H^2}, \quad \Omega_{\Lambda}= \frac{\Lambda}{3 H^2}, \quad \Omega_m=\frac{8 \pi G \rho}{3 H^2}; \quad \quad H=\frac{\dot a(t)}{a(t)} 
\end{equation}
and the Friedmann equations from the other side
\begin{equation}
H^2 = -\frac{k}{a^2(t)}+\frac{\Lambda}{3} + \frac{8 \pi G \rho}{3}\, .
\end{equation}
The Eq.(\ref{RicciL}) becomes
\begin{eqnarray}\label{RicciLP}
\frac{\mathfrak{r}_u}{2}&&= (a_0H_0)^2\left(-\Omega_k +2\Theta\tilde{\delta} \,\Omega_m\right)
\end{eqnarray}
\noindent 
The importance of Eq.(\ref{RicciLP}) is in the fact that, it enables one to study the stability conditions for the sequence of structures i.e. voids and walls with different attributed parameters \cite{GK2}. Then, the Eq.(\ref{gdev}) can be written as
\begin{equation}\label{geolambda}
\frac{d^2 l}{d\lambda^2} +\mathfrak{r}_\Lambda l=0\,, 
\end{equation}
where $d\eta=(a_0H_0)^{-1}d\lambda$, 
$d\lambda \equiv \left(\Omega_\Lambda+(1-\Omega_\Lambda+\Omega_mz)(1+z)^2\right)^{-1/2}dz$.
Now by considering $\Omega_k=0$, Eq.(\ref{geolambda}) becomes
\begin{equation}\label{geodelt}
\frac{d^2 l}{d\tau^2} +\Theta\,\tilde{\delta} l=0, 
\end{equation}
where $\tau=\sqrt{2\Omega_m}\lambda$. 

\section{Variable void sizes}

We assume $\Theta=1$ and for $0=T_0<T_1<T_2<\cdots T_{2i-1}<T_{2i}\cdots<T_{2n-1}<T_{2n}=T$

\begin{equation}\label{period1}
\tilde{\delta}(\tau) =
    \begin{cases}
      -\kappa^2 & T_0< \tau < T_1\\
      +\omega^2 & T_1 < \tau < T_2\\
      \cdots\\
      -\kappa^2 & T_{2i-2}< \tau < T_{2i-1}\\
      +\omega^2 & T_{2i-1} < \tau < T_{2i}\\
      \cdots\\
      -\kappa^2 & T_{2n-2}< \tau < T_{2n-1}\\
      +\omega^2 & T_{2n-1} < \tau < T_{2n}
    \end{cases}    
\end{equation}

We know that

\begin{equation}\label{geodelt}
    x(T)=f^Tx(0)=\left[\prod_{i=1}^n e^{B\tau_{2i}}e^{A\tau_{2i-1}}\right]x(0), 
\end{equation}

where $\tau_k=T_k-T_{k-1}$ for $1\le k\le 2n$
\begin{equation}
A=
\begin{pmatrix}
0 & 1\\
\kappa^2 & 0
\end{pmatrix}\ , \qquad
B=
\begin{pmatrix}
0 & 1\\
-\omega^2 & 0
\end{pmatrix}\ , \qquad
x=
\begin{pmatrix}
\ell\\
\dot{\ell}
\end{pmatrix}\ .
\end{equation}

\begin{equation}
e^{AT_A}=
\begin{pmatrix}
\cosh(\kappa T_A) & \quad\frac{1}{\kappa}\sinh(\kappa T_A)\\
\\
\kappa\sinh(\kappa T_A) & \quad\cosh(\kappa T_A)
\end{pmatrix}\ , \qquad
e^{BT_B}=
\begin{pmatrix}
\cos(\omega T_B) & \quad\frac{1}{\omega}\sin(\omega T_A)\\
\\
-\omega\sin(\omega T_A) & \quad\cos(\omega T_A)
\end{pmatrix}\ ,
\end{equation}
and
\begin{align}
&e^{C(T_A+T_B)}\equiv e^{BT_B}e^{AT_A}\notag\\
&=\left(
\begin{array}{cc}
 \frac{\kappa}{\omega} \sinh (\kappa T_A) \sin (\omega T_B)
 +\cosh (\kappa T_A) \cos (\omega T_B) 
 &\quad \frac{1}{\kappa}\sinh (\kappa T_A) \cos (\omega T_B)
 +\frac{1}{\omega}\cosh (\kappa T_A) \sin (\omega T_B) \\
 \\
 \kappa  \sinh (\kappa T_A) \cos (\omega T_B) 
 -\omega  \cosh (\kappa T_A) \sin (\omega T_B) 
 &\quad \cosh (\kappa T_A) \cos (\omega T_B)
 -\frac{\omega}{\kappa }  \sinh (\kappa T_A) \sin (\omega T_B)
\end{array}\right)\,,
\end{align}
\begin{align}
\text{Det}\left(e^{C(T_A+T_B)}\right)&=1\,,\\
\text{Tr}\left(e^{C(T_A+T_B)}\right)
&=2 \cosh (\kappa T_B) \cos (\omega T_A) +
\left(\frac{\kappa}{\omega } 
-\frac{\omega}{\kappa }\right)  \sinh (\kappa T_B) \sin (\omega T_A)\,.
\end{align}

Our aim is to obtain the distortion parameter $\beta$ for $f^T$. Distortion in this case is defined as follows
\begin{align*}
    \beta=\text{min}(|\lambda_-|,|\lambda_+|)\,,
\end{align*}
where
\begin{equation}
\lambda_\pm=\frac{b\pm\sqrt{b^2-4}}{2}\,,\qquad b=\text{Tr}\left(f^T\right)\,.
\end{equation}
For large $n$ the expression for the trace becomes complicated. For example, when $n=2$, for the trace we get
\begin{eqnarray}\label{Trace}
b=&&\frac{1}{k^2 \omega ^2} \left(\sinh (k \tau _1) \sinh (k \tau _3) [k \omega  (k^2-\omega^2 ) \cosh (k \tau _3) \sin \omega(\tau _2+\tau _4) +\right.\nonumber\\
&&\left.(k^4+\omega ^4) \sin(\tau _2 \omega) \sin(\tau _4 \omega )+2 k^2 \omega ^2 \cos (\tau _2 \omega ) \cos(\tau _4 \omega )\right] +\nonumber\\
&&k\omega \cosh (k \tau _1)\left[2 k \omega  \cosh (k \tau _3) \cos(\tau _2+\tau _4) \omega+\right.\left.\left.(k^2-\omega^2 ) \sinh (k \tau _3) \sin (\tau _2+\tau _4) \omega \right]\right).
\end{eqnarray}

First, let us consider two voids and two walls, i.e. when $\tau_1=\tau_3$, $\tau_2=\tau_4$, $\kappa\tau_1\ll 1$ and $\omega\tau_2\ll 1$. Then we get
  \begin{eqnarray}{\label{dist1a}}
 \beta=1 -T \sqrt{-\left<\delta_0\right>}\,,
 \end{eqnarray}
 where
 \begin{equation}
     T=2(\tau_1+\tau_2)\,,\qquad \left<\delta_0\right> = \frac{-\kappa^2\tau_1+\omega^2\tau_2}{\tau_1+\tau_2}\,.
 \end{equation}
 
To describe the variation of the void size over the redshift, we repeat the same procedure for the case when the next void is slightly larger than the former one, i.e. when $\tau_3$ is slightly larger than $\tau_1$, $\tau_3=\tau_1+\epsilon$, $\epsilon\ll\tau_1$, and for the distortion we get
\begin{eqnarray}
 \beta=1
 -T \sqrt{-\left<\delta_0\right>} +\frac{\delta_{void}-\left<\delta_0\right>}{\sqrt{-\left<\delta_0\right>}}\epsilon\,.
 \end{eqnarray}
 
For the case of $n$ pairs of voids and walls as described above, one can easily get
\begin{align}
    b&=\text{Tr}\left(\prod_{i=1}^n e^{B\tau_{2}}e^{A(\tau_{1}+\epsilon_i)}\right)
    =\text{Tr}\left(\prod_{i=1}^n e^{B\tau_{2}}e^{A\tau_{1}}e^{A\epsilon_i}\right)
    =\text{Tr}\left(\prod_{i=1}^n e^{B\tau_{2}}e^{A\tau_{1}}(I+A\epsilon_i)\right)+o(\epsilon)\nonumber\\
    &=\text{Tr}\left(\prod_{i=1}^n e^{B\tau_{2}}e^{A\tau_{1}}\right)
    +\text{Tr}\left(\prod_{i=1}^n e^{B\tau_{2}}e^{A\tau_{1}}A\right)\epsilon+o(\epsilon)\nonumber\\
    &=\text{Tr}\left(\left[e^{B\tau_{2}}e^{A\tau_{1}}\right]^n\right)
    +\text{Tr}\left(\left[e^{B\tau_{2}}e^{A\tau_{1}}\right]^nA\right)\epsilon+o(\epsilon)
\end{align}
where $\epsilon=\sum_{i=1}^n\epsilon_i$ and $n(\tau_1+\tau_2)+\epsilon = T$.

\begin{align}
    \beta
    &=1
    -n \sqrt{-(\tau_{1}+\tau_{2}) \left(-\kappa ^2 \tau_{1} +\omega ^2\tau_{2} \right)}\nonumber\\
    &\quad-\tfrac{1}{2}\left(\frac{1}{\sqrt{-(\tau_{1}+\tau_{2}) 
    \left(-\kappa ^2 \tau_{1} +\omega ^2\tau_{2} \right)}}     
    -n \right)\left(\kappa ^2 (\tau_{2} +\tau_{1})-\left(-\kappa ^2 \tau_{1} +\tau_{2} \omega ^2\right)\right)\epsilon +o(\epsilon)\nonumber\\
    &=1 -n(\tau_1+\tau_2)\sqrt{-\left<\delta_0\right>}
    -\left(1 -n(\tau_{2} +\tau_{1})\sqrt{-\langle\delta_0\rangle} \right)
    \frac{\left(\delta_{\text{void}} -\left<\delta_0\right>\right)}{2\sqrt{-\langle\delta_0\rangle}}\epsilon +o(\epsilon)\nonumber\\
    &=\left[1 -n(\tau_1+\tau_2)\sqrt{-\left<\delta_0\right>}\right]\left[
    1-
    \frac{\left(\delta_{\text{void}} -\left<\delta_0\right>\right)}{2\sqrt{-\langle\delta_0\rangle}}\epsilon\right] +o(\epsilon)
\end{align}
  
Figs.1 and 2 illustrate these formulae for given ratio of the void and wall sizes and $\epsilon$, the size increase parameter of the voids.

\begin{figure}[h]
\caption{The photon beam distortion due to the void hyperbolicity for given void and wall size ratios and void size variation $\epsilon=0.1$.}
\centering
\includegraphics[]{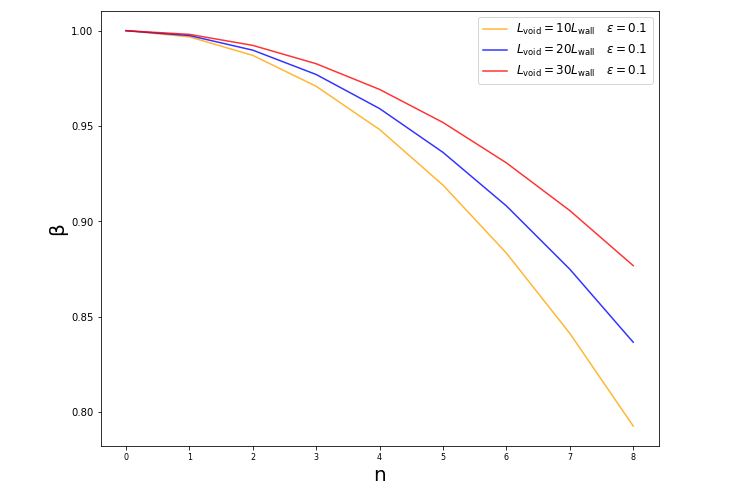}
\end{figure}

\begin{figure}[h]
\caption{The same as in Fig.1 but for fixed void and wall size ratio and $\epsilon= 0.1, 0.05, 0.02$.}
\centering
\includegraphics[]{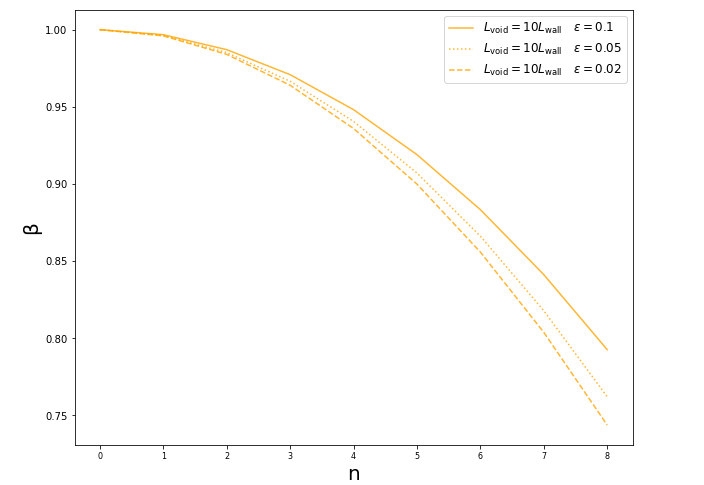}
\end{figure}

\section{Conclusions}

The GIGANTES dataset \cite{Kr} marks the appearance of precision cosmology landscape from cosmic voids. Such extensive data enable
to formulate and study to higher precision theoretical problems associated to redshift dependence of the voids' mean physical parameters. Particularly, \cite{Kr} provide the median radius distribution of the voids up to redshift $z=1$, which can be directly compared with the predictions of the theory of formation of the cosmic web, i.e. of the sequence of semi-periodical underdense regions, the voids, separated by thin, overdensity walls.

We studied the hyperbolicity properties of the voids, i.e. their ability for deviation the photon beams, so that void's underdensity has the effective role of the negative curvature in the Jacobi equation. Developing further our previous studies \cite{Sam1,Sam2}, we now revealed the cumulative distortion of the images due to the sequence of line-of-sight voids in the case of varying sizes of the voids, i.e. as is indicated by the survey \cite{Kr}. The formula for the distortion parameter is obtained, quantitatively illustrated in Figs. 1 and 2, for given void/wall size ratio and the increase in the size of the voids in the sequence. 

The tangential image distortion in the galactic surveys are detected long ago (e.g. \cite{Pea,Guz,Per}), typically reflecting the contribution of redshift (Doppler) effects such as e.g. the Kayser effect, and to which (to the distortion) the hyperbolicity of voids has to contribute as well.    

The redshift dependence of the voids' parameters carries crucial cosmological information. Along with that, the redshift dependence of the void sizes has to influence the observational effects associated with the voids at high redshifts, as revealed in this study obtaining a formula for the quantitative evaluation of the hyperbolicity by voids of variable sizes at cosmological distances.     
  
M.S. acknowledges the support of Armenian SCS grant 20RF-142.

%\section{Acknowledgment}

\end{document}